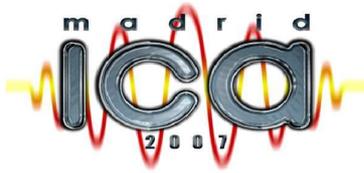

**19th INTERNATIONAL CONGRESS ON ACOUSTICS
MADRID, 2-7 SEPTEMBER 2007**

# Experimental validation and physical modelling of vocal folds pathologies

Ruty, Nicolas[1]; Brutel-Vuilmet, Claire[1]; Pelorson, Xavier[1]; Van Hirtum, Annemie[1]
[1]Gipsa-Lab, Département Parole et Cognition; 46 Avenue Félix Viallet, 38031 Grenoble France;
nicolas.ruty@gipsa-lab.inpg.fr

**ABSTRACT**
Voiced sounds involve self-sustained vocal folds oscillations due to the interaction between the airflow and the vocal folds. Common vocal folds pathologies like polyps and anatomical asymmetry degrade the mechanical vocal fold properties and consequently disturb the normal oscillation pattern resulting in an abnormal sound production. Treatment of voice abnormalities would benefit from an improved understanding between the pathology and the resulting oscillation pattern which motivates physical vocal folds modelling. The current study applies a theoretical vocal folds model to vocal folds pathologies. The theoretical vocal folds model is validated using an experimental set-up simulating the human phonatory apparatus. It consists in a pressure reservoir, a self-oscillating latex replica of the vocal folds and an acoustical resonator. The effects of pathologies are simulated by modifying the replica's geometry, elasticity, and homogeneity under controlled experimental conditions. In general, we observed a close match between measurements and theoretical predictions, which is all the more surprising considering the crudeness of the theoretical model.

**INTRODUCTION**
Voiced sound production results from the interaction between the airflow coming from the lungs and deformable walls (vocal folds). Under certain condition, self sustained oscillations of the vocal folds appear. Some pathology (polyps, paralysis, anatomical asymmetry of the vocal folds) disturb this oscillation. Consequently, sound production becomes difficult and sometimes impossible. More precisely, paralysis of a vocal fold, total or partial, can result in bitonal sound production. As observed by Mergell et al. [1] the two vocal folds oscillate independently and the damaged vocal fold oscillates at a higher frequency.
Distributed vocal fold models (mass-spring models), coupled to a simple description of the airflow and the vocal tract acoustic, are often used [2]. The aim of this study is to test if this type of model is able to reproduce pathological behaviors of the vocal folds.
The theoretical model, taking into account the effects of asymmetry, is briefly described. Next, results of simulation are compared to experimental data. The data are acquired from measurements performed on a experimental set-up, which is based on a asymmetrical vocal folds replica.

**THEORY**
The outlined theoretical larynx model is derived from the one presented in [3]. It consists in a quasi-steady flow model taking into account viscous effects, a mechanical vocal fold model for which parameters can be independently fixed in each vocal fold, and an acoustical propagation model based on linear acoustic. This theoretical model, schematized in figure 1, is analyzed with two different methods (stability analysis, dynamical analysis).

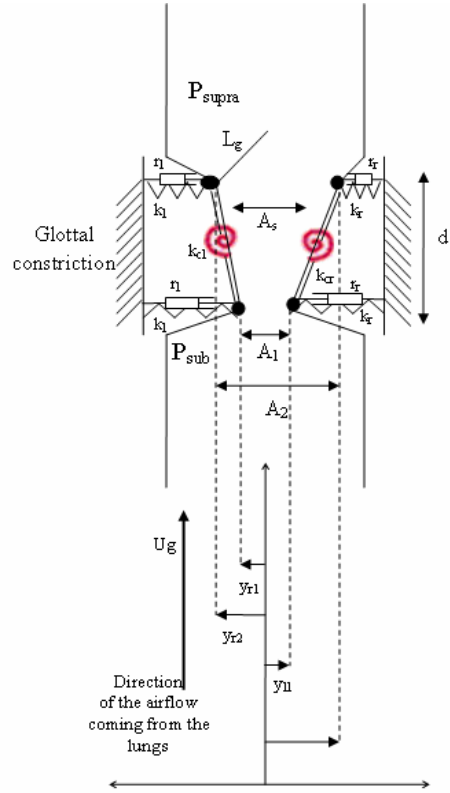

Figure 1.- Schematic overview of the asymmetric vocal folds model with $P_{sub}$ the subglottal pressure, $P_{supra}$ the supraglottal pressure, $L_g$ the glottal width, d glottis length. $A_1$, $A_2$, $A_s$ are the section areas at each masse and at airflow separation point. $y_{l1}$, $y_{l2}$, $y_{r1}$, $y_{r2}$ are the mass positions compared to the central axis. $k_l$ and $k_r$ are the spring stiffnesses. $k_{cl}$ and $k_{cr}$ are the coupling spring stiffnesses. $r_l$ and $r_r$ are the damping constants.

**Fluid flow description**
In pre-phonatory condition, the vocal folds constitute a constriction of which geometry is depicted in figure 1. The airflow coming from the lungs is modeled as incompressible, and quasi-steady. Viscous effects are taken into account by a Poiseuille corrective term. Airflow is separating from the glottis at a point $x_s$ at which energy is dissipated by turbulence effects. The pressure distribution along the glottis constriction can be described by the following equation:

$$P(x,t) = P_{sub} - \frac{1}{2}\rho U_g^2 \left(\frac{1}{A^2(x,t)}\right) - 12\mu L_g^2 U_g \int_{x_0}^{x} \frac{dx}{A^3(x,t)}, \text{ si } x < x_s \quad (Eq.1)$$

$$P(x,t) = P_{supra}, \text{ si } x > x_s,$$

where $P_{sub}$ and $P_{supra}$ are the sub and supra glottal pressures, $\rho$ the air density, $U_g$ the glottal flow, $\mu$ the dynamic viscosity coefficient, $L_g$ the glottal width, $A(x,t)$ the section area for a given x-coordinate and at time t.

The glottal flow is calculated by taking into account the pressure drop between the entrance of the glottis and the airflow separation point. Pressure forces acting on the vocal folds are determined using this airflow model.

**Mechanical description**
Each vocal fold is represented by two coupled oscillators. It consists in two masses linked to the body and coupled by springs ($k_l$, $k_r$, $k_c$) and dampers ($r_l$, $r_r$). Interactions with the airflow are taken into account using the pressure forces. Equation 2 describes the movements of the left vocal fold and equation 3 the movements of the right vocal fold.

$$\begin{cases} \frac{m_l}{2}\frac{\partial^2 y_{l1}(t)}{\partial t^2} = -k_l(y_{l1}(t) - y_{l10}) - k_{cl}(y_{l1}(t) - y_{l10} - y_{l2}(t) + y_{l20}) - r_l\frac{\partial y_{l1}(t)}{\partial t} + F_{l1} \\ \frac{m_l}{2}\frac{\partial^2 y_{l2}(t)}{\partial t^2} = -k_l(y_{l2}(t) - y_{l20}) - k_{cl}(y_{l2}(t) - y_{l20} - y_{l1}(t) + y_{l10}) - r_l\frac{\partial y_{l2}(t)}{\partial t} + F_{l2} \end{cases} \quad (Eq. 2)$$

$$\begin{cases} \dfrac{m_r}{2}\dfrac{\partial^2 y_{r1}(t)}{\partial t^2} = -k_r(y_{r1}(t)-y_{r10}) - k_{cr}(y_{r1}(t)-y_{r10}-y_{r2}(t)+y_{r20}) - r_r\dfrac{\partial y_{r1}(t)}{\partial t} + F_{r1} \\ \dfrac{m_r}{2}\dfrac{\partial^2 y_{r2}(t)}{\partial t} = -k_r(y_{r2}(t)-y_{r20}) - k_{cr}(y_{r2}(t)-y_{r20}-y_{r1}(t)+y_{r10}) - r_r\dfrac{\partial y_{r2}(t)}{\partial t} + F_{r2} \end{cases} \quad \text{(Eq. 3)}$$

where $m_i$ (i={l,r}) represents the vibrating mass of a vocal fold, $y_{i1}$, $y_{i2}$ are the mass positions relative to the central axe, $y_{i10}$, $y_{i20}$ the rest positions, $k_l$ and $k_r$ the spring stiffnesses, $k_{cl}$ and $k_{cr}$ the coupling spring stiffnesses and $r_l$, $r_r$ the damping constants.

Asymmetry can be introduced in two different ways. It can be either a geometrical asymmetry if the rest positions $y_{l10}$ and $y_{l20}$ are different from $y_{r10}$ and $y_{r20}$ or a mechanical asymmetry if the set of parameters $k_c$, k, r and m are not the same for both vocal folds. In the literature, asymmetry is quantified by an asymmetry factor Q such as $k_r = Qk_l \; et \; m_r = m_l/Q$ ([4], [1]). In this study, we quantify asymmetry between the two folds (geometrical and mechanical) based on frequency responses experimentally measured on the vocal fold replica.

**Acoustical modelling**
The vocal tract is modeled by a uniform resonator (section S=2.5cm², length L=17cm). The glottis, due to its dimensions and the frequencies band of interest (50-400Hz), is considered like an acoustical point source. Impedance at the entrance of the resonator is calculated in the following equation,

$$Z = p/u = Z_0 \frac{Z_L \cos(kL) + iZ_0 \sin(kL)}{iZ_L \sin(kL) + Z_0 \cos(kL)} \quad \text{(Eq. 4)}$$

where p is the acoustic pressure, u is the acoustic flow, $Z_0 = \rho c/S$, c the speed of sound, $Z_L = Z_0\left((k.a)^2/2 + i\frac{8}{3\pi}(k.a)\right)$, a the resonator radius, $k = \dfrac{2\pi.f}{\left(c.(1-1.65\times 10^{-3}/(a\sqrt{f}))\right)} - \dfrac{3i\times 10^{-5}\sqrt{f}}{a}$, f the frequency.

The first resonance of this impedance is extracted. It results in equation 5 :

$$\frac{\partial^2 \psi(t)}{\partial t^2} + \frac{\omega_A}{Q_A}\frac{\partial \psi(t)}{\partial t} + \omega_A^2 \psi(t) = \frac{Z_A \omega_A}{Q_A} u \quad \text{(Eq. 5)}$$

with $p = \dfrac{\partial \psi}{\partial t}$, $\omega_a$ the resonance pulsation, $Q_a$ the quality factor of this resonance, and $Z_a$ the amplitude.

**Analysis**
We use two methods for studying this theoretical model. First, a stability analysis can be performed. Equations are linearised around and equilibrium state. For a given set of control parameters (mechanical, aerodynamical and acoustical parameters), the eigenvalues of the system give access to two pertinent values in speech: subglottal pressure threshold of oscillation and fundamental oscillation frequencies. Then, it is also interesting to observe the temporal evolution of the characteristic values of the system, and to compare it directly with measurements performed on the experimental set-up. The described set of equations need to be discretized ([2]).

**EXPERIMENTAL SET-UP**

**Description**
As shown in figure 2, this experimental set-up is composed of a pressure reservoir ("the lungs") fed by a compressor. A vocal fold replica is fixed on it. This replica is made of metal pieces covered with latex ([i], [j]) and filled with water under pressure ([d] et [d']). Water pressure in each "vocal fold" is controlled independently. These two "vocal folds" are pressed between two metallic pieces to guarantee sealing. A uniform section pipe ("the vocal tract") of 17 cm length is fixed to the replica. Airflow is forced through the replica and self-sustained oscillations appear.

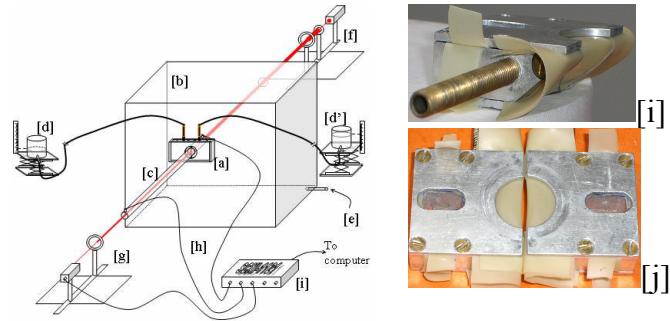

Figure 2.- Experimental set-up. [a] vocal fold replica. [b] pressure reservoir (lungs). [c] acoustic pipe of 17 cm. [d], [d'] water pressure reservoirs. [e] air admission. [f,g] laser diode, optical lenses and photodiode. [h] pressure sensors. [i] signal conditioner.

Asymmetrical conditions can be studied with the presented replica imposing a different water pressure in each "vocal fold". It corresponds to "pathological conditions". A modification of this pressure corresponds to a modification of the mechanical and geometrical characteristics of the replica. The influence of the asymmetrical condition on the """"%%% quantity can be experimentally assessed.

Thanks to pressure sensors, we measure the pressure upstream the replica ("subglottal pressure"), the pressure downstream the replica ("supraglottal pressure") and the pressure at the end of the resonator ("acoustic pressure at lips"). The variations of the constriction aperture are measured with an optical system. The measured data allow to extract the quantities which are predicted with the theoretical model, the oscillation threshold, fundamental frequency, dynamical evolution.

**Link between experimental set-up and theoretical model**
The relationship between the vocal fold replica and the theoretical model is not trivial for all the control parameters. Some of them, like the subglottal pressure, the initial geometry or the acoustical characteristics are directly connected to reality (upstream pressure of the "latex vocal fold", aperture measured by the optical system, length and section of the downstream resonator). Some parameters have to be estimated (masses, spring stiffnesses, damping constants). The vibrating mass is estimated as $m_{cv}=m_{water}/2$, the quantity of water contained in the latex vocal fold replica divided by a factor 2. Spring stiffnesses and damping constant are connected to the mechanical resonance of the replica as follows, $\omega_0 = \sqrt{2k/m}$, $Q_0 = m\omega_0/2r$ where $\omega_0$ is the resonance pulsation and $Q_0$ the quality factor of this mechanical resonance.

The mechanical responses are measured as described by [3]. For a given internal water pressure in the replica, the relationship with a set of control parameters of the theoretical model can be established.

**RESULTS**

**Oscillation pressure thresholds and fundamental frequencies**
With symmetrical initial condition as a reference, we have studied the effect of an increasing asymmetry. Experimentally, the internal water pressure in the left "vocal fold" is fixed at 3000 Pa while the water pressure in the right "vocal fold" varied between 3000 Pa and 8000 Pa by step of 500 Pa. For each condition, the upstream pressure is increased until oscillations appear. Upstream pressure is stabilized in order to get steady condition, and next the pressure is decreased until oscillation ùùùù%%%. The different oscillations modes and pressure threshold are analyzed. By a stability analysis, oscillation pressure threshold and fundamental frequency of the theoretical model are calculated. Results are presented in figure 3. In the asymmetrical case, a first oscillation mode appears for a small value of the upstream pressure (~ 400 Pa). A second oscillation mode appears for a higher pressure value. One can observe that asymmetry causes an increase of the pressure threshold, corresponding to a difficult oscillation production.

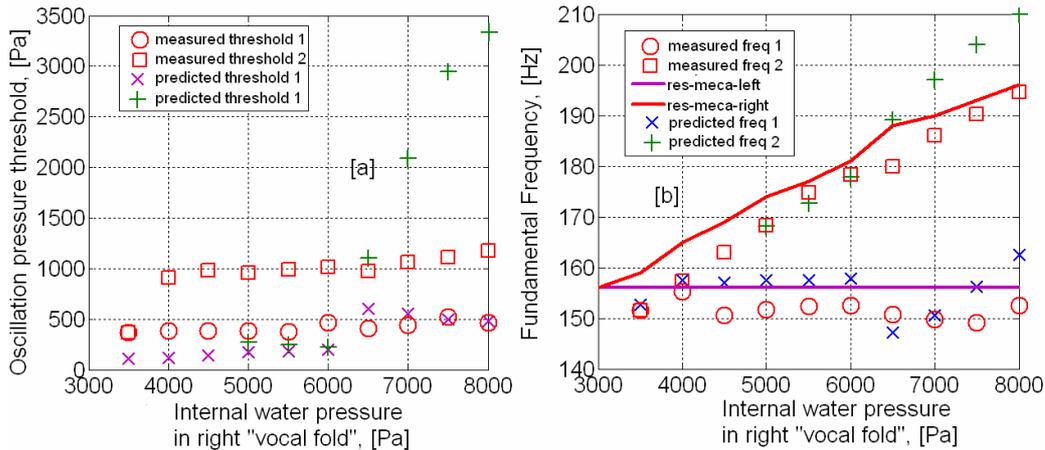

Figure 3.- Comparison between measured data ant theoretical predictions. [a] oscillation pressure thresholds. [b] oscillation fundamental frequency

These phenomena are reproduced by the theoretical model, which predicts two oscillation modes and an increase of oscillation pressure threshold, corresponding to a forcing. Despite of this qualitative agreement, these oscillation onset pressures are underestimated by the theoretical predictions, but the orders of magnitude remain coherent with what is observed experimentally.

Oscillation frequencies are well predicted. Two oscillation modes are observed, which corresponds to the mechanical resonance of each "vocal fold". The low frequency is linked to the left "vocal fold" (~150 Hz, res-meca-left). The high frequency is linked to the right "vocal fold" for which resonance frequency increases with an increase in internal water pressure (res-meca-right).

**Dynamics**

Experimental data can also be analyzed from a temporal point of view. For a given internal water pressure in each "vocal fold", control parameters (geometry and mechanics) are set into the theoretical model. A dynamical simulation is performed and the theoretical prediction can be compared to experimental data as shown in figure 4.

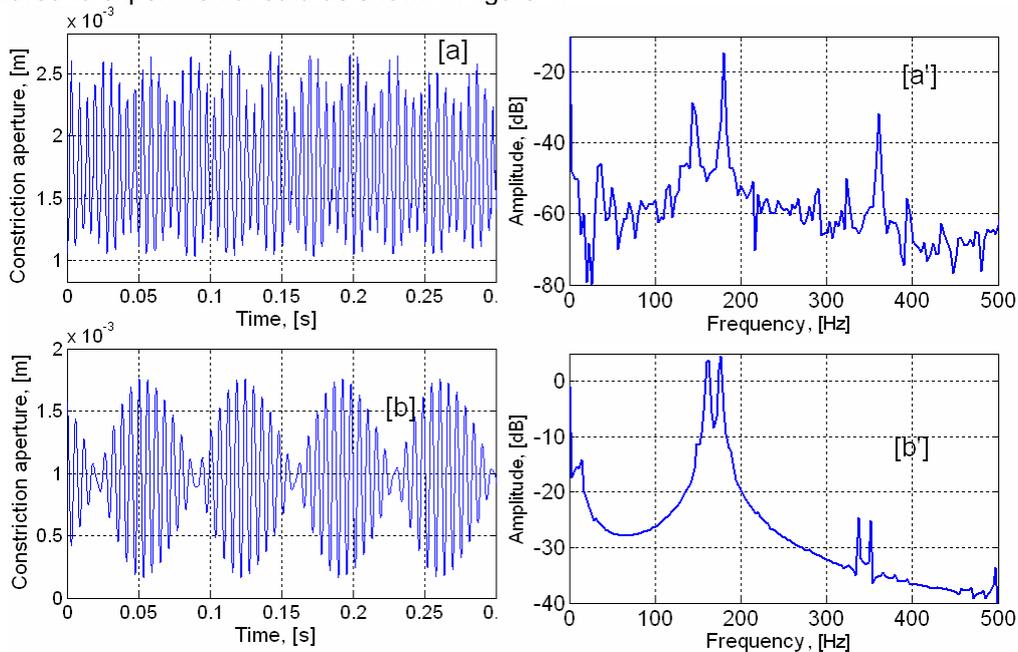

Figure 4.- Comparison between experimental data and theoretical predictions. [a] measurement of the aperture between the two latex vocal folds.[a'] spectrum of this signal. [b] theoretical prediction of the aperture between the two vocal fold for a given set of mechanical and geometrical parameters, estimated using the mechanical response. [b'] spectrum of this signal.

In this simulation, the water pressure is set to 3000 Pa in the left "vocal fold" and to 6000 Pa in the right "vocal fold". Experimental data show two oscillation frequencies, as is also observed « in-vivo » by [1]. In the spectrum, two close peaks are observed. Each one corresponds two the mechanical resonance of each "vocal fold". The theoretical model has a very similar behavior. The modulations and the two peaks in the spectrum are predicted. Amplitude of the simulated oscillations are close to the one observed experimentally. The predicted equilibrium position is lower than the experimental equilibrium positions.

**CONCLUSION**
A theoretical vocal fold model has been confronted to experimental data obtained on an experimental set-up. The experimental conditions were set to reproduce a unilateral disease of a vocal fold. Our conclusions are as follows:
- Important phenomenon like the apparition of multiphonics (bitonality) is well predicted by the theoretical model. Frequencies are predicted with a good accuracy.
- Vocal forcing (need to increase the lung pressure) in presence of asymmetry is predicted but only in a qualitative way. Oscillation pressure threshold are overestimated in case of strong asymmetry.
- In case of dynamical simulation, the predicted amplitude of vocal folds oscillations approximate to experimental data.
According to the simplicity of the theoretical model, these results are particularly interesting. Some modifications, in order to take into account non linear phenomena, are subject to further research in order to improve the theoretical model. One can note that such models are only useful for global pathologies. Lesions or local disease of vocal folds like polyps can't be taken into account. Continuous models (membrane, finite element model) are necessary for such studies.